\documentclass[prl,letterpaper,aps,10pt,twocolumn,floatfix,showpacs]{revtex4-2}

\usepackage{graphicx}
\usepackage{dcolumn}
\usepackage{bm}
\usepackage{color,soul}
\usepackage{amsmath}
\usepackage{upgreek}
\usepackage{siunitx,upgreek}

\begin{document}

\title{Hybrid Nonlinear Effects in Photonic Integrated Circuits}
\author{Arghadeep \surname{Pal}$^{\;1,2,\dagger}$}
\author{Alekhya \surname{Ghosh$^{\;1,2,\dagger}$}}
\author{Shuangyou \surname{Zhang}$^{\;3,1,*}$}
\author{Toby \surname{Bi}$^{\;1,2}$}
\author{Masoud \surname{Kheyri}$^{\;1,2}$}
\author{Haochen \surname{Yan}$^{\;1,2}$}
\author{Yaojing \surname{Zhang}$^{\;4}$}
\author{Pascal \surname{Del'Haye$^{\;1,2,*}$}}
\affiliation{$^1$Max Planck Institute for the Science of Light, Staudtstra{\ss}e 2,
D-91058 Erlangen, Germany\\$^2$Department of Physics, Friedrich Alexander University Erlangen-Nuremberg, D-91058 Erlangen, Germany\\$^3$Department of Electrical and Photonics Engineering, Technical University of Denmark, Kgs. Lyngby, 2800, Denmark\\$^4$School of Science and Engineering, The Chinese University of Hong Kong (Shenzhen), Guangdong, 518172, P.R. China}
\affiliation{$^\dagger$These authors contributed equally}
\affiliation{$^*$Corresponding authors: shzhan@dtu.dk, pascal.delhaye@mpl.mpg.de}

\begin{abstract}
Nonlinear optics in photonic integrated circuits is usually limited to utilizing the nonlinearity of a single material.
In this work, we demonstrate the use of hybrid optical nonlinearities that occur in two different materials. This approach allows us to observe combined Raman scattering and Kerr frequency comb generation using silicon nitride (Si$_3$N$_4$) microresonators with fused silica cladding. Here, the fused silica cladding provides Raman gain, while the Si$_3$N$_4$ core provides the Kerr nonlinearity for frequency comb generation. This way we can add Raman scattering to an integrated photonic silicon nitride platform, in which Raman scattering has not been observed so far because of insufficient Raman gain. The Raman lasing is observed in the silica-clad Si$_3$N$_4$ resonators at an on-chip optical power of 143 mW, which agrees with theoretical simulations. This can be reduced to mw-level with improved optical quality factor. Broadband Raman-Kerr frequency comb generation is realized through dispersion engineering of the waveguides. The use of hybrid optical nonlinearities in multiple materials opens up new functionalities for integrated photonic devices, e.g. by combining second and third-order nonlinear materials for combined supercontinuum generation and self-referencing of frequency combs. Combining materials with low threshold powers for different nonlinearities can be the key to highly efficient nonlinear photonic circuits for compact laser sources, high-resolution spectroscopy, frequency synthesis in the infrared and UV, telecommunications and quantum information processing.  
\end{abstract}

\maketitle
Most research on nonlinear integrated microresonators exploits the nonlinearities in the core of the waveguide, which has a higher refractive index compared to the cladding. 
In these devices, the intensity of the light that is confined in the core imposes a constraint on the attainable nonlinearities.
In addition, the choice of nonlinear optical materials needs to be tailored to specific applications, depending on the required nonlinear process.
Enforcing light to interact with multiple materials during propagation helps to exploit multiple optical nonlinearities simultaneously with interesting implications for the fields of 
meta-materials~\cite{liu2016hybrid}, plasmonics~\cite{kauranen2012nonlinear} and photonic crystals~\cite{joannopoulos1997photonic}. 
Hybrid photonic integration has been very successful, e.g. for 
developing on-chip lasers~\cite{guo2022chip, liu2024fully} and integrating materials with different optical properties~\cite{terrasanta2022aluminum}. However, combining multiple 
nonlinear light-matter interactions in different materials has not yet been observed in integrated microresonators. In this work, we demonstrate a hybrid nonlinear microresonator that combines Raman scattering in a fused silica cladding with Kerr-nonlinearity induced frequency comb generation in a Si$_3$N$_4$ core.
\\
\indent Stimulated Raman scattering (SRS) is a nonlinear optical phenomenon that arises from the interaction between light and molecular vibrations of the host medium. The Raman effect generates light at new frequencies 
and plays a pivotal role in a wide range of applications such as spectroscopy~\cite{smith2019modern,han2021surface}, optical communications~\cite{spillane2002ultralow,ahmadi2021widely,semrau2017achievable}, and quantum information processing~\cite{reim2011single}. Initial works on Raman lasing required high pump power by pulse excitation~\cite{stolen1984development}. Recent advances in nanofabrication of optical microresonators with ultra-high quality (\textit{Q}) factors 
provide a fertile ground for observing such nonlinear light interactions at extremely low power due to their small mode volumes
~\cite{kippenberg2011microresonator}. Raman lasing in microresonators has been reported in silica~\cite{spillane2002ultralow,min2005controlled,kato2017transverse,suzuki2018broadband}, silicon~\cite{griffith2016coherent, zhang2022broadband, zhang2021raman}, aluminum nitride~\cite{liu2017integrated, liu2018integrated}, lithium niobate~\cite{yu2020raman}, silicon carbide~\cite{li2024efficient} and others~\cite{hausmann2014diamond,choi2019low,xia2022engineered}.
\\
\indent Si$_3$N$_4$ has emerged as a promising platform for integrated photonics over the past decade~\cite{okawachi2011octave,brasch2016photonic,rebolledo2022coherent,rahim2017expanding,zhang2024low,chiles2018deuterated,yaojing_SSB} due to ultralow propagation loss, a relatively high Kerr nonlinearity, and an ultra-broad transparency window, which ensure negligible multi-photon absorption in the telecom band. However, Raman lasing or Raman comb generation has not been observed in Si$_3$N$_4$ integrated photonics due to its relatively low Raman gain~\cite{karpov2016raman}.
\begin{figure*}
\includegraphics[width=2\columnwidth]{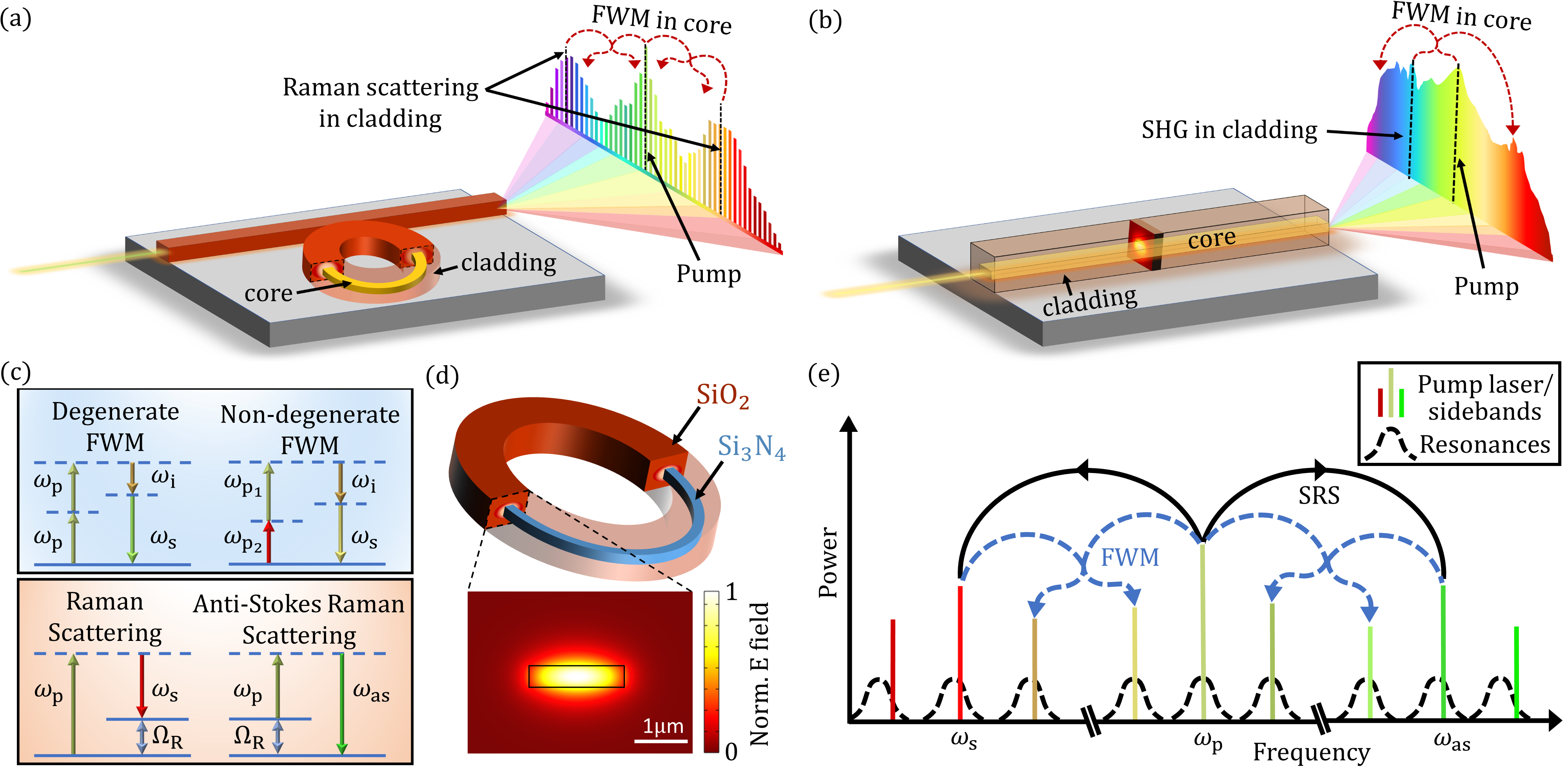}
\caption {\textit{Hybrid nonlinear interactions in different materials in photonic integrated circuits.
} (a) A single frequency continuous wave pump laser initiates Raman scattering in the cladding of the ring resonator, which yields a broad frequency comb via four-wave mixing (FWM) in the core material. (b) Proposed model to generate supercontinuum in integrated systems utilizing the hybrid nonlinearities of the core and cladding materials. The second harmonic generation (SHG) in the cladding and FWM in the core material combine to generate supercontinuum.
(c) Energy diagram showing the different FWM processes and Raman scattering. For FWM, $\omega_\text{p}$, $\omega_{\text{p}_1}$, and $\omega_{\text{p}_2}$ represent the pump frequencies, with $\omega_\text{s}$ and $\omega_\text{i}$ being the signal and the idler frequencies, respectively. In the case of Raman scattering, $\omega_{\text{p}}$, $\omega_\text{s}$, $\omega_{\text{as}}$, and $\Omega_\text{R}$ are the frequencies of the pump, Stokes, anti-Stokes, and the Raman shift. (d) 3D illustration of the optical mode in a microresonator with Si$_3$N$_4$ core and silica (SiO$_2$) cladding. The mode profile (normalised Electric field) is shown for a Si$_3$N$_4$ microresonator with 1.8$~\upmu$m width and 0.4$~\upmu$m height of the core. (e) Schematic of the hybrid Raman-scattering and four-wave mixing process. The Stokes and anti-Stokes signals generated from the pump at a frequency spacing of $\Omega_\text{R}$ are indicated by black arrows, whereas the sidebands around the pump, Stokes, and anti-stokes signals generated by FWM are highlighted by blue dotted arrows. The solid lines of different colors indicate the pump laser and the generated equidistant comb lines. The cavity modes are affected by the normal dispersion of the resonator and are shown via the black dashed lines.}
\label{fig1}
\end{figure*}
\begin{figure*}[t]
\includegraphics[width=2\columnwidth]{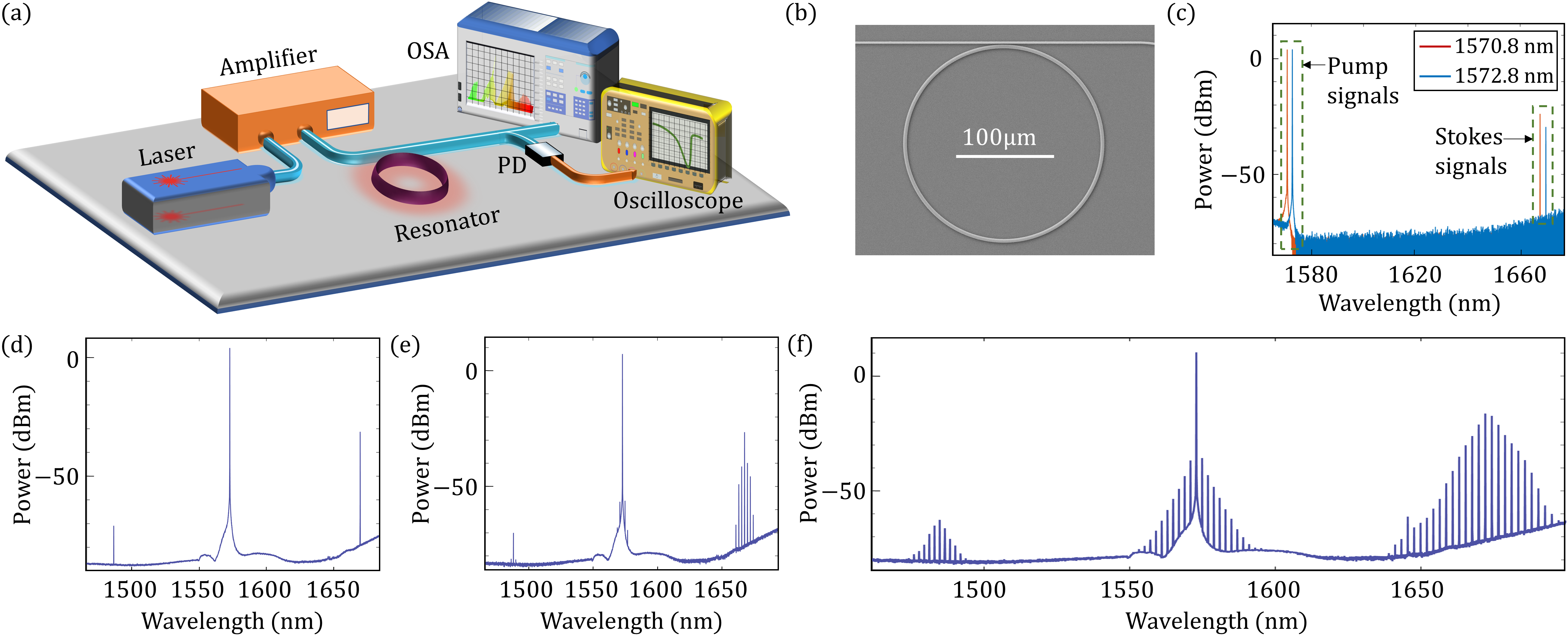}
\caption {\textit{Experimental setup and the transition from Raman lasing to hybrid Raman-Kerr comb generation.} (a) Experimental setup for Raman lasing and Raman-Kerr comb generation. OSA: Optical spectrum analyzer, PD: Photodiode. (b) Scanning electron microscope (SEM) image of the $100~\upmu$m radius Si$_{3}$N$_{4}$ microresonator with $1.8~\upmu$m width and $400$~nm height. (c) A superimposed spectrum showing different single-mode Raman lasing peaks for its corresponding pump wavelengths, with the Raman shift being $11~$THz in both cases. (d) Appearance of the Stokes ($1669.76~$nm) and the anti-Stokes sidebands ($1486.48~$nm) around the pump wavelength ($1572.8~$nm) with a Raman shift of $11~$GHz. (e) Four-wave-mixing induced comb formation around the Stokes wavelength. (f) Raman-Kerr comb lines generate around the pump, Stokes, and anti-Stokes wavelengths.}
\label{fig2}
\end{figure*}
\\ \indent In this work, we exploit optical nonlinearities in the core and the cladding of a resonator to demonstrate Raman lasing and the subsequent generation of Kerr frequency combs in integrated Si$_3$N$_4$ microresonators. By engineering the geometry of Si$_3$N$_4$ microresonators, the optical field overlap with the silica cladding is optimized to effectively stimulate the silica Raman effect. Coupled with the high four-wave mixing (FWM) gain of the Si$_3$N$_4$ core, this enables the generation of a broadband Kerr comb through a Raman-scattering assisted process. We begin by experimentally demonstrating Raman lasing in Si$_3$N$_4$ microresonators, along with broadband comb generation spanning the pump, Stokes, and anti-Stokes wavelength regions. Next, we vary the waveguide width to engineer resonator dispersion for broadband Kerr frequency comb generation. Theoretical calculations reveal the optimal Si$_3$N$_4$ waveguide thickness to achieve low-threshold Raman lasing within the silica cladding.
\\ \indent This novel method enables control over nonlinear frequency conversion through the integration of cladding materials with diverse nonlinear properties, departing from the conventional approach where nonlinearity is confined to a single core material. We believe that this demonstration can enable many new applications across the visible and mid-infrared spectral regions.
\\
\indent Figure~\ref{fig1}(a,b) show examples of emerging nonlinear phenomena utilizing proposed hybrid nonlinearities in different materials. Figure~\ref{fig1}(c) highlights the energy diagram for the different FWM and Raman scattering processes. Signal ($\omega_\text{s}$) and idler ($\omega_\text{i}$) photons are generated (as shown in Fig.~\ref{fig1}(c)) from two similar pump photons ($\omega_\text{p}$) in case of degenerate FWM, whereas for non-degenerate FWM, the two pump photons have different frequencies ($\omega_{\text{p}_1}$, $\omega_{\text{p}_2}$). Figure~\ref{fig1}(c) also shows Raman scattering where the pump photon ($\omega_\text{p}$) transfers energy to the scattering molecule (Raman shift $\Omega_\text{R}$), thereby generating a lower frequency Stokes photon ($\omega_\text{s}$). For anti-Stokes-Raman scattering, energy from the scattering molecule and the pump is transferred to a higher frequency anti-Stokes ($\omega_\text{as}$) photon. Figure~\ref{fig1}(d) shows the 3D-structure of a microresonator with core and cladding, together with a cross section of the mode profile at a wavelength of 
around $1570~$nm. 
The optical field is partially outside of the core and thus, the structure facilitates the interaction of the optical field with the cladding material. The generation of Raman-scattering assisted Kerr combs occurs in two stages: first, light interaction with the silica cladding produces a strong Raman signal ($\omega_\text{s}$). This Raman signal then propagates through the Si$_3$N$_4$ core, where it serves as a secondary pump, generating comb sidebands through nondegenerate FWM with the primary pump ($\omega_\text{p}$).
Figure~\ref{fig1}(e) shows an illustration of the generation of Raman-scattering assisted Kerr combs in an integrated Si$_3$N$_4$ chip. The solid black arrows and the blue dashed arrows indicate the FWM and the SRS processes, respectively. These two nonlinearities generate further comb lines shown in different colors in Fig.~\ref{fig1}(e).
\\ \indent We start by exploring the Raman lasing in Si$_{3}$N$_{4}$ ring resonators. Figure~\ref{fig2}(a) shows our experimental setup with a continuous-wave external cavity laser being tuned into the resonance from the blue detuned side. The laser is amplified using an erbium-doped fiber amplifier and coupled to the ring resonator using a lensed fiber. 
\\ \indent We fabricate the Si$_3$N$_4$ microresonators using commercial 400-nm-thick Si$_3$N$_4$ wafers with a 3~$\upmu$m silica box layer. After the Si$_{3}$N$_{4}$ etching process, another 3-$\upmu$m-thickness silica layer is deposited as the top cladding layer. Details on the fabrication process can be found in Ref.~\cite{zhang2024low, zhang2023geometry}. Figure~\ref{fig2}(b) shows a scanning electron microscope (SEM) image of a 100-$\upmu$m-radius microresonator (free spectral range, FSR 241 GHz) with a core width of 1.8~$\upmu$m and a height of 400~nm. The fundamental mode (TE$_{00}$) has an intrinsic \textit{Q}-factor of 2.2 million measured at 1570~nm pump wavelength (shown in Appendix~A). Both the fundamental (TE$_{00}$) mode and higher order (TE$_{10}$) mode are in normal dispersion regime (Appendix~A).
\\ \indent Pumping the resonator at $1570.8$~nm yields single-mode Raman lasing at $1667.5$~nm as shown in Fig.~\ref{fig2}(c). Shifting the pump by one FSR (at $1572.8~$nm) causes an equidistant shift in the Raman lasing wavelength ($1669.7$~nm), thereby maintaining the Raman shift of $11$~THz. This verifies that the generated sideband originates from Raman scattering instead of mode crossing assisted comb generation in the normal dispersion regime, where the mode-crossing pins down the frequency of the first sideband.
\\ \indent Figure~\ref{fig2}(d) shows the nonlinear SRS process occurring in the microresonator with both the Stokes (at $1669.76$~nm) and the anti-Stokes (at $1486.48$~nm) sidebands, each being $11$ THz away from the input pump (at $1572.8$~nm). The microresonator is in a normal dispersion regime, where similar to bichromatic pumping of resonators~\cite{strekalov2009generation,zhang2020spectral,zhang2022dark}, frequency comb lines can be generated from the interaction of the pump and the Stokes sideband. With an increase in the input power (beyond $150~$mW), the nondegenerate FWM process induces sidebands around the pump, Stokes, and anti-Stokes line as shown in Fig.~\ref{fig2}(e). Further increase in pump power results in the formation of broadband comb spectra, as shown in Fig.~\ref{fig2}(f).
\\ \indent Dispersion leads to a change in the FSR at different wavelengths. This FSR mismatch around the pump and the Stokes sideband 
hinders the comb generation~\cite{zhang2020spectral,zhang2022dark}. However, FSR matching at two pump wavelengths can drive broadband comb generation. In the following section, we focus on optimizing the microresonator geometry to shift this FSR matching point closer to the Raman Stokes signal, enhancing comb generation efficiency and bandwidth.
\break
\par \noindent \textbf{Broadband Raman comb via dispersion engineering} -- 
\begin{figure}[b]
\includegraphics[width=1\columnwidth]{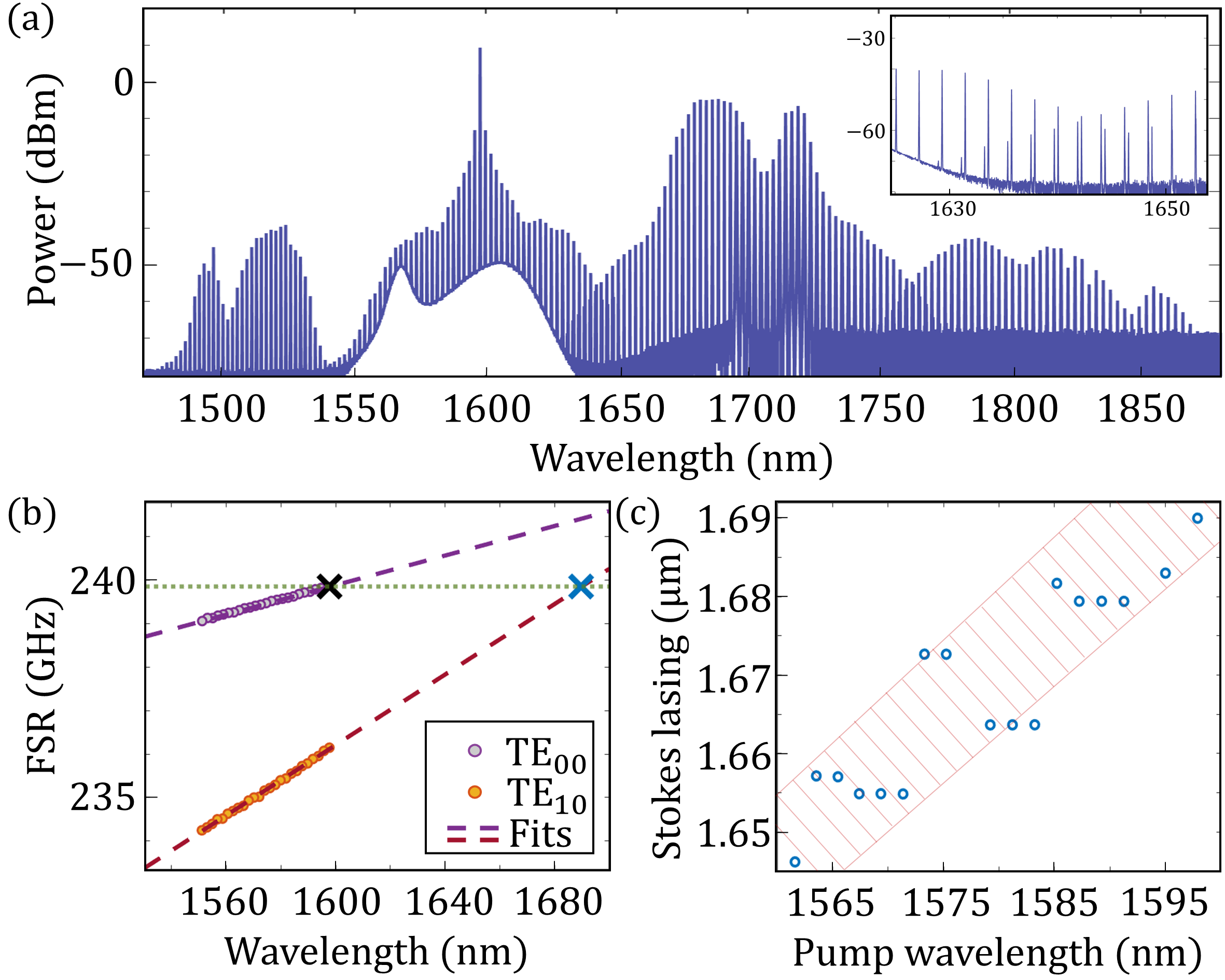}
\caption {\textit{Raman comb generation in a dispersion engineered microresonator with 1.9~$\mu$m core-width.} (a) Cascaded Raman comb generation around the pump, the anti-Stokes, and the first-order and second-order Stokes wavelength spanning over 400~nm generated via pumping at 1598~nm. The inset shows the crossing of the comb lines coming from the two different mode families. (b) Measured FSR data of TE$_{00}$ (violet circles) and TE$_{10}$ (red circles) resonant mode families (dashed lines show linear curve fittings). The FSR value of TE$_{00}$ mode at the pump wavelength (at 1598~nm, marked by black cross) matches the FSR of TE$_{10}$ mode at 1689.8~nm (blue cross). (c) The shift in the Stokes lasing (blue circles) with respect to the pump wavelength. The red dashed region represents the boundary of the data points, thereby highlighting the trend in the variation of Raman lasing with the pump.} 
\label{fig3}
\end{figure}
By adjusting the Si$_3$N$_{4}$ waveguide width to 1.9~\text{$\upmu$}m, we observe cascaded Raman-Kerr comb formation as shown in Fig.~\ref{fig3}(a), while pumping at $1598$~nm. The comb lines are generated even around the second-order Raman Stokes wavelength along with the anti-Stokes, Stokes, and pump wavelengths, spanning over 400~nm. The inset shows the spectral region where the comb around the pump and the comb around the first Stokes sideband overlap, revealing that the two combs originate from two different mode families. The number of comb lines around the Stokes sideband in microresonators with $1.9~\upmu$m wide waveguides doubles compared to Si$_{3}$N$_{4}$ rings with $1.8~\upmu$m core width. 
The reason for triggering this broadband comb formation is revealed in the changes in FSR for both the TE$_{00}$ and TE$_{10}$ modes, as illustrated in Fig.~\ref{fig3}(b). The FSR of the TE$_{00}$ mode at the pump wavelength aligns with that of the TE$_{10}$ mode at 1689.8~nm, which is close to the Raman lasing wavelength. Due to the limited wavelength range of the input laser (1510~nm to 1630~nm), we extrapolate the measured data using linear fits, indicated by the dashed lines. The Raman lasing depends on the phase matching between the Stokes sideband within the Raman gain spectrum and the intracavity field at the pump wavelength. Figure~\ref{fig3}(c) shows the wavelength shift of the single-mode Raman lasing with the shift in input pump wavelength for the resonator with engineered dispersion. The red dashed region marks the range of the Stokes wavelength's variations and highlights a linear trend in the Stokes shift with respect to the pump wavelength. The graph (Fig.~\ref{fig3}(c)) shows that pumping the resonator at around $1598$~nm leads to Stokes lasing (at $1690$~nm, shown in Fig.~\ref{fig3}(b)) near the FSR matching point.\\

\par \noindent \textbf{Raman lasing threshold in Si$_{3}$N$_{4}$ resonator} --
The threshold power ($P_{\mathrm{t_R}}$) required to generate Raman lasing in microresonators can be expressed as~\cite{kippenberg2004theoretical}:
\begin{equation}
P_{\mathrm{t_R}} = C\left(\Gamma\right)\frac{\pi^2n^2}{g_{\text{R}}\ \lambda_{\text{P}}\lambda_{\text{R}} P_\text{clad}}V_{\text{eff}}\left(\frac{1}{\textit{Q}_\text{0}}\right)^2\frac{\left(1+K\right)^3}{K}.
\label{LLEquations}
\end{equation}
Here, $C(\Gamma)$ denotes the correlation factor accounting for the intermodal coupling between the counterpropagating fields (in this work we consider no intermodal coupling, $C(\Gamma) = 1$), $n$ is the refractive index of the medium with $\lambda_\text{P}$ and $\lambda_\text{R}$ being the pump and Raman wavelengths, respectively. Since the Raman lasing is generated in the fused silica cladding, the effective mode volume ($V_{\text{eff}}$) contributing to the SRS process is calculated only in the cladding region using the finite element method (FEM). P$_{\text{clad}}$ is the percentage of input power interacting with the cladding medium. The nonlinear bulk Raman gain coefficient is $g_{\text{R}}$ and the normalized coupling parameter is given by $K$. Assuming the quality factors and coupling factors at the Stokes and pump wavelengths being equal, the minimum threshold occurs at $K = 0.5$~\cite{kippenberg2004theoretical}. 
\begin{figure}[b]
\includegraphics[width=1\columnwidth]{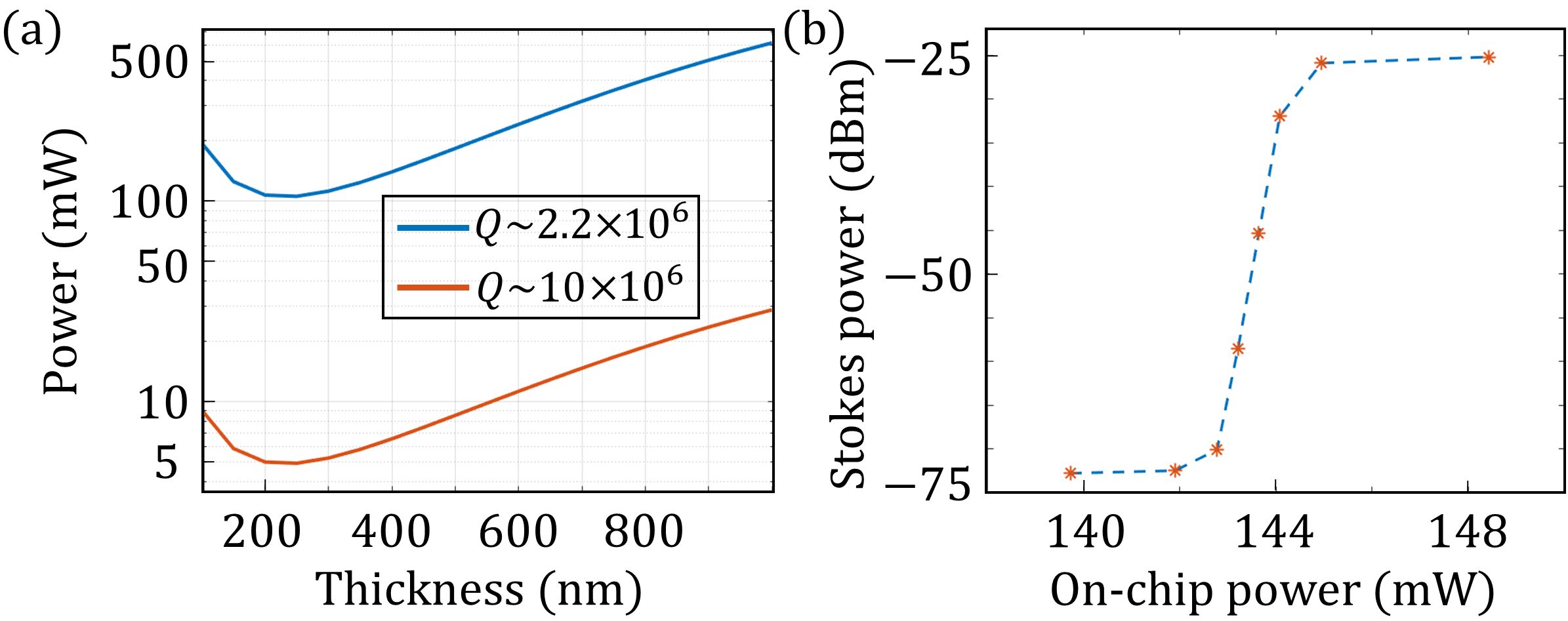}
\caption {\textit{Comparison between theoretically calculated and experimentally obtained Raman lasing threshold.} Panel (a) shows calculated Raman lasing threshold power as a function of the Si${_3}$N${_4}$ core height for Q-factors of $2.2$ million (in blue) and $10$ million (in red), keeping other parameters fixed ($100~\upmu$m radius, $1.8~\upmu$m width). (b) Stokes sideband power as a function of the on-chip power showing the experimentally observed Raman lasing threshold to be $143~$mW.}
\label{fig4}
\end{figure}
\indent Figure~\ref{fig4}(a) shows the variation in the calculated threshold power required for Raman lasing in a silica cladded Si$_{3}$N$_{4}$ resonator with changes in the resonator's thickness, at a constant core-width and radius of $1.8~\upmu$m and $100~\upmu$m, respectively. For a microresonator with a \textit{Q} factor of 2.2 million (blue line), the required threshold power exhibits a minimum of 106 mW when the core (made of Si$_3$N$_4$) is $250$~nm thick. The threshold power increases with a change of the thickness in either direction. 
The threshold power for Raman lasing can be further reduced to a few mW by enhancing the quality factor to 10 million (red line) through optimized fabrication techniques. 
\\ \indent By plotting the measured power of the generated Stokes sideband as a function of the input power in Fig.~\ref{fig4}(b), we obtain a threshold input power of $143~$mW for Raman lasing, which is close to our theoretically calculated threshold (Fig.~\ref{fig4}(a)). This agreement between the theoretical and experimental results further confirms that the Raman lasing is driven by the field interactions with silica in the cladding, in addition to the 11 THz Raman shift, which is characteristic for fused silica.
\\ \par \noindent \textbf{Discussion and outlook} -- In summary, we demonstrate hybrid optical nonlinearities in two separate nonlinear materials in photonic integrated circuits. In our experiments, we demonstrate this by combining Kerr frequency comb generation in the core of a Si$_3$N$_4$ microresonator with Raman scattering in the fused silica cladding. By tailoring the thickness of the Si$_3$N$_4$ core, we optimize the optical mode overlap within the cladding material to exploit high Raman gain in silica and high FWM gain in Si$_3$N$_4$. Theoretical simulations reveal the optimum Si$_3$N$_4$ core thickness for low threshold Raman lasing. Our experiments demonstrate a controlled transition from the Raman Stokes-anti-Stokes sideband pair generation to the formation of frequency combs. Moreover, we adjust the resonator width to engineer its dispersion and increase the comb bandwidths around the Stokes and anti-Stokes sidebands. The hybridization of multiple optical nonlinearities in different materials can be of high interest for advanced photonic integrated circuits. This approach allows to combine ideal materials for each required nonlinearity in the system. Another potential application is the combination of materials with high second-order and high third-order nonlinearities for improved supercontinuum generation together with second harmonic generation for self-referencing of frequency combs.
\\
\par \noindent \textbf{Acknowledgements} -- This work was funded by the European Union’s H2020 ERC Starting Grant ``CounterLight” 756966, the Max Planck Society, and the Max Planck School of Photonics. SZ acknowledge support from Deutsche Forschungsgemeinschaft project 541267874.\\
\indent AP and AG contributed equally to this work. AP and AG did the experiments and theoretical calculations. SZ did the fabrication. SZ and PD supervised the project. All authors discussed the results.\\\\


\bibliography{Refs}

\begin{thebibliography}{42}%
\makeatletter
\providecommand \@ifxundefined [1]{%
 \@ifx{#1\undefined}
}%
\providecommand \@ifnum [1]{%
 \ifnum #1\expandafter \@firstoftwo
 \else \expandafter \@secondoftwo
 \fi
}%
\providecommand \@ifx [1]{%
 \ifx #1\expandafter \@firstoftwo
 \else \expandafter \@secondoftwo
 \fi
}%
\providecommand \natexlab [1]{#1}%
\providecommand \enquote  [1]{``#1''}%
\providecommand \bibnamefont  [1]{#1}%
\providecommand \bibfnamefont [1]{#1}%
\providecommand \citenamefont [1]{#1}%
\providecommand \href@noop [0]{\@secondoftwo}%
\providecommand \href [0]{\begingroup \@sanitize@url \@href}%
\providecommand \@href[1]{\@@startlink{#1}\@@href}%
\providecommand \@@href[1]{\endgroup#1\@@endlink}%
\providecommand \@sanitize@url [0]{\catcode `\\12\catcode `\$12\catcode `\&12\catcode `\#12\catcode `\^12\catcode `\_12\catcode `\%12\relax}%
\providecommand \@@startlink[1]{}%
\providecommand \@@endlink[0]{}%
\providecommand \url  [0]{\begingroup\@sanitize@url \@url }%
\providecommand \@url [1]{\endgroup\@href {#1}{\urlprefix }}%
\providecommand \urlprefix  [0]{URL }%
\providecommand \Eprint [0]{\href }%
\providecommand \doibase [0]{http://dx.doi.org/}%
\providecommand \selectlanguage [0]{\@gobble}%
\providecommand \bibinfo  [0]{\@secondoftwo}%
\providecommand \bibfield  [0]{\@secondoftwo}%
\providecommand \translation [1]{[#1]}%
\providecommand \BibitemOpen [0]{}%
\providecommand \bibitemStop [0]{}%
\providecommand \bibitemNoStop [0]{.\EOS\space}%
\providecommand \EOS [0]{\spacefactor3000\relax}%
\providecommand \BibitemShut  [1]{\csname bibitem#1\endcsname}%
\let\auto@bib@innerbib\@empty
\bibitem [{\citenamefont {Liu} \emph {et~al.}(2016)\citenamefont {Liu}, \citenamefont {Kang}, \citenamefont {Mayer}, and \citenamefont {Werner}}]{liu2016hybrid}%
  \BibitemOpen
  \bibfield  {author} {\bibinfo {author} {\bibfnamefont {L.}~\bibnamefont {Liu}}, \bibinfo {author} {\bibfnamefont {L.}~\bibnamefont {Kang}}, \bibinfo {author} {\bibfnamefont {T.~S.} \bibnamefont {Mayer}},  and \bibinfo {author} {\bibfnamefont {D.~H.} \bibnamefont {Werner}}, }\bibfield  {title} {\enquote {\bibinfo {title} {{Hybrid metamaterials for electrically triggered multifunctional control}},} }\href@noop {} {\bibfield  {journal} {\bibinfo  {journal} {Nat. Commun.} }\textbf {\bibinfo {volume} {7}}, \bibinfo {pages} {13236} (\bibinfo {year} {2016})}\BibitemShut {NoStop}%
\bibitem [{\citenamefont {Kauranen} and \citenamefont {Zayats}(2012)}]{kauranen2012nonlinear}%
  \BibitemOpen
  \bibfield  {author} {\bibinfo {author} {\bibfnamefont {M.}~\bibnamefont {Kauranen}} and \bibinfo {author} {\bibfnamefont {A.~V.} \bibnamefont {Zayats}}, }\bibfield  {title} {\enquote {\bibinfo {title} {{Nonlinear plasmonics}},} }\href@noop {} {\bibfield  {journal} {\bibinfo  {journal} {Nat. Photon.} }\textbf {\bibinfo {volume} {6}}, \bibinfo {pages} {737--748} (\bibinfo {year} {2012})}\BibitemShut {NoStop}%
\bibitem [{\citenamefont {Joannopoulos} \emph {et~al.}(1997)\citenamefont {Joannopoulos}, \citenamefont {Villeneuve}, and \citenamefont {Fan}}]{joannopoulos1997photonic}%
  \BibitemOpen
  \bibfield  {author} {\bibinfo {author} {\bibfnamefont {J.~D.} \bibnamefont {Joannopoulos}}, \bibinfo {author} {\bibfnamefont {P.~R.} \bibnamefont {Villeneuve}},  and \bibinfo {author} {\bibfnamefont {S.}~\bibnamefont {Fan}}, }\bibfield  {title} {\enquote {\bibinfo {title} {{Photonic crystals}},} }\href@noop {} {\bibfield  {journal} {\bibinfo  {journal} {Solid State Commun.} }\textbf {\bibinfo {volume} {102}}, \bibinfo {pages} {165--173} (\bibinfo {year} {1997})}\BibitemShut {NoStop}%
\bibitem [{\citenamefont {Guo} \emph {et~al.}(2022)\citenamefont {Guo}, \citenamefont {McLemore}, \citenamefont {Xiang}, \citenamefont {Lee}, \citenamefont {Wu}, \citenamefont {Jin}, \citenamefont {Kelleher}, \citenamefont {Jin}, \citenamefont {Mason}, \citenamefont {Chang} \emph {et~al.}}]{guo2022chip}%
  \BibitemOpen
  \bibfield  {author} {\bibinfo {author} {\bibfnamefont {J.}~\bibnamefont {Guo}}, \bibinfo {author} {\bibfnamefont {C.~A.} \bibnamefont {McLemore}}, \bibinfo {author} {\bibfnamefont {C.}~\bibnamefont {Xiang}}, \bibinfo {author} {\bibfnamefont {D.}~\bibnamefont {Lee}}, \bibinfo {author} {\bibfnamefont {L.}~\bibnamefont {Wu}}, \bibinfo {author} {\bibfnamefont {W.}~\bibnamefont {Jin}}, \bibinfo {author} {\bibfnamefont {M.}~\bibnamefont {Kelleher}}, \bibinfo {author} {\bibfnamefont {N.}~\bibnamefont {Jin}}, \bibinfo {author} {\bibfnamefont {D.}~\bibnamefont {Mason}}, \bibinfo {author} {\bibfnamefont {L.}~\bibnamefont {Chang}},  \emph {et~al.}, }\bibfield  {title} {\enquote {\bibinfo {title} {{Chip-based laser with 1-hertz integrated linewidth}},} }\href@noop {} {\bibfield  {journal} {\bibinfo  {journal} {Sci. Adv.} }\textbf {\bibinfo {volume} {8}}, \bibinfo {pages} {eabp9006} (\bibinfo {year} {2022})}\BibitemShut {NoStop}%
\bibitem [{\citenamefont {Liu} \emph {et~al.}(2024)\citenamefont {Liu}, \citenamefont {Qiu}, \citenamefont {Ji}, \citenamefont {Bancora}, \citenamefont {Lihachev}, \citenamefont {Riemensberger}, \citenamefont {Wang}, \citenamefont {Voloshin}, and \citenamefont {Kippenberg}}]{liu2024fully}%
  \BibitemOpen
  \bibfield  {author} {\bibinfo {author} {\bibfnamefont {Y.}~\bibnamefont {Liu}}, \bibinfo {author} {\bibfnamefont {Z.}~\bibnamefont {Qiu}}, \bibinfo {author} {\bibfnamefont {X.}~\bibnamefont {Ji}}, \bibinfo {author} {\bibfnamefont {A.}~\bibnamefont {Bancora}}, \bibinfo {author} {\bibfnamefont {G.}~\bibnamefont {Lihachev}}, \bibinfo {author} {\bibfnamefont {J.}~\bibnamefont {Riemensberger}}, \bibinfo {author} {\bibfnamefont {R.~N.} \bibnamefont {Wang}}, \bibinfo {author} {\bibfnamefont {A.}~\bibnamefont {Voloshin}},  and \bibinfo {author} {\bibfnamefont {T.~J.} \bibnamefont {Kippenberg}}, }\bibfield  {title} {\enquote {\bibinfo {title} {{A fully hybrid integrated Erbium-based laser}},} }\href@noop {} {\bibfield  {journal} {\bibinfo  {journal} {Nat. Photon.} , \bibinfo {pages} {1--7}} (\bibinfo {year} {2024})}\BibitemShut {NoStop}%
\bibitem [{\citenamefont {Terrasanta} \emph {et~al.}(2022)\citenamefont {Terrasanta}, \citenamefont {Sommer}, \citenamefont {M{\"u}ller}, \citenamefont {Althammer}, \citenamefont {Gross}, and \citenamefont {Poot}}]{terrasanta2022aluminum}%
  \BibitemOpen
  \bibfield  {author} {\bibinfo {author} {\bibfnamefont {G.}~\bibnamefont {Terrasanta}}, \bibinfo {author} {\bibfnamefont {T.}~\bibnamefont {Sommer}}, \bibinfo {author} {\bibfnamefont {M.}~\bibnamefont {M{\"u}ller}}, \bibinfo {author} {\bibfnamefont {M.}~\bibnamefont {Althammer}}, \bibinfo {author} {\bibfnamefont {R.}~\bibnamefont {Gross}},  and \bibinfo {author} {\bibfnamefont {M.}~\bibnamefont {Poot}}, }\bibfield  {title} {\enquote {\bibinfo {title} {{Aluminum nitride integration on silicon nitride photonic circuits: a hybrid approach towards on-chip nonlinear optics}},} }\href@noop {} {\bibfield  {journal} {\bibinfo  {journal} {Opt. Express} }\textbf {\bibinfo {volume} {30}}, \bibinfo {pages} {8537--8549} (\bibinfo {year} {2022})}\BibitemShut {NoStop}%
\bibitem [{\citenamefont {Smith} and \citenamefont {Dent}(2019)}]{smith2019modern}%
  \BibitemOpen
  \bibfield  {author} {\bibinfo {author} {\bibfnamefont {E.}~\bibnamefont {Smith}} and \bibinfo {author} {\bibfnamefont {G.}~\bibnamefont {Dent}}, }\href@noop {} {\emph {\bibinfo {title} {{Modern Raman spectroscopy: a practical approach}}}} (\bibinfo  {publisher} {John Wiley \& Sons}, \bibinfo {year} {2019})\BibitemShut {NoStop}%
\bibitem [{\citenamefont {Han} \emph {et~al.}(2021)\citenamefont {Han}, \citenamefont {Rodriguez}, \citenamefont {Haynes}, \citenamefont {Ozaki}, and \citenamefont {Zhao}}]{han2021surface}%
  \BibitemOpen
  \bibfield  {author} {\bibinfo {author} {\bibfnamefont {X.~X.} \bibnamefont {Han}}, \bibinfo {author} {\bibfnamefont {R.~S.} \bibnamefont {Rodriguez}}, \bibinfo {author} {\bibfnamefont {C.~L.} \bibnamefont {Haynes}}, \bibinfo {author} {\bibfnamefont {Y.}~\bibnamefont {Ozaki}},  and \bibinfo {author} {\bibfnamefont {B.}~\bibnamefont {Zhao}}, }\bibfield  {title} {\enquote {\bibinfo {title} {{Surface-enhanced Raman spectroscopy}},} }\href@noop {} {\bibfield  {journal} {\bibinfo  {journal} {Nat. Rev. Methods Primers} }\textbf {\bibinfo {volume} {1}}, \bibinfo {pages} {87} (\bibinfo {year} {2021})}\BibitemShut {NoStop}%
\bibitem [{\citenamefont {Spillane} \emph {et~al.}(2002)\citenamefont {Spillane}, \citenamefont {Kippenberg}, and \citenamefont {Vahala}}]{spillane2002ultralow}%
  \BibitemOpen
  \bibfield  {author} {\bibinfo {author} {\bibfnamefont {S.}~\bibnamefont {Spillane}}, \bibinfo {author} {\bibfnamefont {T.}~\bibnamefont {Kippenberg}},  and \bibinfo {author} {\bibfnamefont {K.}~\bibnamefont {Vahala}}, }\bibfield  {title} {\enquote {\bibinfo {title} {{Ultralow-threshold Raman laser using a spherical dielectric microcavity}},} }\href@noop {} {\bibfield  {journal} {\bibinfo  {journal} {Nature} }\textbf {\bibinfo {volume} {415}}, \bibinfo {pages} {621--623} (\bibinfo {year} {2002})}\BibitemShut {NoStop}%
\bibitem [{\citenamefont {Ahmadi} \emph {et~al.}(2021)\citenamefont {Ahmadi}, \citenamefont {Shi}, and \citenamefont {LaRochelle}}]{ahmadi2021widely}%
  \BibitemOpen
  \bibfield  {author} {\bibinfo {author} {\bibfnamefont {M.}~\bibnamefont {Ahmadi}}, \bibinfo {author} {\bibfnamefont {W.}~\bibnamefont {Shi}},  and \bibinfo {author} {\bibfnamefont {S.}~\bibnamefont {LaRochelle}}, }\bibfield  {title} {\enquote {\bibinfo {title} {{Widely tunable silicon Raman laser}},} }\href@noop {} {\bibfield  {journal} {\bibinfo  {journal} {Optica} }\textbf {\bibinfo {volume} {8}}, \bibinfo {pages} {804--810} (\bibinfo {year} {2021})}\BibitemShut {NoStop}%
\bibitem [{\citenamefont {Semrau} \emph {et~al.}(2017)\citenamefont {Semrau}, \citenamefont {Killey}, and \citenamefont {Bayvel}}]{semrau2017achievable}%
  \BibitemOpen
  \bibfield  {author} {\bibinfo {author} {\bibfnamefont {D.}~\bibnamefont {Semrau}}, \bibinfo {author} {\bibfnamefont {R.}~\bibnamefont {Killey}},  and \bibinfo {author} {\bibfnamefont {P.}~\bibnamefont {Bayvel}}, }\bibfield  {title} {\enquote {\bibinfo {title} {{Achievable rate degradation of ultra-wideband coherent fiber communication systems due to stimulated Raman scattering}},} }\href@noop {} {\bibfield  {journal} {\bibinfo  {journal} {Opt. Express} }\textbf {\bibinfo {volume} {25}}, \bibinfo {pages} {13024--13034} (\bibinfo {year} {2017})}\BibitemShut {NoStop}%
\bibitem [{\citenamefont {Reim} \emph {et~al.}(2011)\citenamefont {Reim}, \citenamefont {Michelberger}, \citenamefont {Lee}, \citenamefont {Nunn}, \citenamefont {Langford}, and \citenamefont {Walmsley}}]{reim2011single}%
  \BibitemOpen
  \bibfield  {author} {\bibinfo {author} {\bibfnamefont {K.}~\bibnamefont {Reim}}, \bibinfo {author} {\bibfnamefont {P.}~\bibnamefont {Michelberger}}, \bibinfo {author} {\bibfnamefont {K.}~\bibnamefont {Lee}}, \bibinfo {author} {\bibfnamefont {J.}~\bibnamefont {Nunn}}, \bibinfo {author} {\bibfnamefont {N.}~\bibnamefont {Langford}},  and \bibinfo {author} {\bibfnamefont {I.}~\bibnamefont {Walmsley}}, }\bibfield  {title} {\enquote {\bibinfo {title} {{Single-photon-level quantum memory at room temperature}},} }\href@noop {} {\bibfield  {journal} {\bibinfo  {journal} {Phys. Rev. Lett.} }\textbf {\bibinfo {volume} {107}}, \bibinfo {pages} {053603} (\bibinfo {year} {2011})}\BibitemShut {NoStop}%
\bibitem [{\citenamefont {Stolen} \emph {et~al.}(1984)\citenamefont {Stolen}, \citenamefont {Lee}, and \citenamefont {Jain}}]{stolen1984development}%
  \BibitemOpen
  \bibfield  {author} {\bibinfo {author} {\bibfnamefont {R.~H.} \bibnamefont {Stolen}}, \bibinfo {author} {\bibfnamefont {C.}~\bibnamefont {Lee}},  and \bibinfo {author} {\bibfnamefont {R.}~\bibnamefont {Jain}}, }\bibfield  {title} {\enquote {\bibinfo {title} {{Development of the stimulated Raman spectrum in single-mode silica fibers}},} }\href@noop {} {\bibfield  {journal} {\bibinfo  {journal} {J. Opt. Soc. Am. B} }\textbf {\bibinfo {volume} {1}}, \bibinfo {pages} {652--657} (\bibinfo {year} {1984})}\BibitemShut {NoStop}%
\bibitem [{\citenamefont {Kippenberg} \emph {et~al.}(2011)\citenamefont {Kippenberg}, \citenamefont {Holzwarth}, and \citenamefont {Diddams}}]{kippenberg2011microresonator}%
  \BibitemOpen
  \bibfield  {author} {\bibinfo {author} {\bibfnamefont {T.~J.} \bibnamefont {Kippenberg}}, \bibinfo {author} {\bibfnamefont {R.}~\bibnamefont {Holzwarth}},  and \bibinfo {author} {\bibfnamefont {S.~A.} \bibnamefont {Diddams}}, }\bibfield  {title} {\enquote {\bibinfo {title} {{Microresonator-based optical frequency combs}},} }\href@noop {} {\bibfield  {journal} {\bibinfo  {journal} {Science} }\textbf {\bibinfo {volume} {332}}, \bibinfo {pages} {555--559} (\bibinfo {year} {2011})}\BibitemShut {NoStop}%
\bibitem [{\citenamefont {Min} \emph {et~al.}(2005)\citenamefont {Min}, \citenamefont {Yang}, and \citenamefont {Vahala}}]{min2005controlled}%
  \BibitemOpen
  \bibfield  {author} {\bibinfo {author} {\bibfnamefont {B.}~\bibnamefont {Min}}, \bibinfo {author} {\bibfnamefont {L.}~\bibnamefont {Yang}},  and \bibinfo {author} {\bibfnamefont {K.}~\bibnamefont {Vahala}}, }\bibfield  {title} {\enquote {\bibinfo {title} {{Controlled transition between parametric and Raman oscillations in ultrahigh-Q silica toroidal microcavities}},} }\href@noop {} {\bibfield  {journal} {\bibinfo  {journal} {Appl. Phys. Lett.} }\textbf {\bibinfo {volume} {87}} (\bibinfo {year} {2005})}\BibitemShut {NoStop}%
\bibitem [{\citenamefont {Kato} \emph {et~al.}(2017)\citenamefont {Kato}, \citenamefont {Hori}, \citenamefont {Suzuki}, \citenamefont {Fujii}, \citenamefont {Kobatake}, and \citenamefont {Tanabe}}]{kato2017transverse}%
  \BibitemOpen
  \bibfield  {author} {\bibinfo {author} {\bibfnamefont {T.}~\bibnamefont {Kato}}, \bibinfo {author} {\bibfnamefont {A.}~\bibnamefont {Hori}}, \bibinfo {author} {\bibfnamefont {R.}~\bibnamefont {Suzuki}}, \bibinfo {author} {\bibfnamefont {S.}~\bibnamefont {Fujii}}, \bibinfo {author} {\bibfnamefont {T.}~\bibnamefont {Kobatake}},  and \bibinfo {author} {\bibfnamefont {T.}~\bibnamefont {Tanabe}}, }\bibfield  {title} {\enquote {\bibinfo {title} {{Transverse mode interaction via stimulated Raman scattering comb in a silica microcavity}},} }\href@noop {} {\bibfield  {journal} {\bibinfo  {journal} {Opt. Express} }\textbf {\bibinfo {volume} {25}}, \bibinfo {pages} {857--866} (\bibinfo {year} {2017})}\BibitemShut {NoStop}%
\bibitem [{\citenamefont {Suzuki} \emph {et~al.}(2018)\citenamefont {Suzuki}, \citenamefont {Kubota}, \citenamefont {Hori}, \citenamefont {Fujii}, and \citenamefont {Tanabe}}]{suzuki2018broadband}%
  \BibitemOpen
  \bibfield  {author} {\bibinfo {author} {\bibfnamefont {R.}~\bibnamefont {Suzuki}}, \bibinfo {author} {\bibfnamefont {A.}~\bibnamefont {Kubota}}, \bibinfo {author} {\bibfnamefont {A.}~\bibnamefont {Hori}}, \bibinfo {author} {\bibfnamefont {S.}~\bibnamefont {Fujii}},  and \bibinfo {author} {\bibfnamefont {T.}~\bibnamefont {Tanabe}}, }\bibfield  {title} {\enquote {\bibinfo {title} {{Broadband gain induced Raman comb formation in a silica microresonator}},} }\href@noop {} {\bibfield  {journal} {\bibinfo  {journal} {J. Opt. Soc. Am. B} }\textbf {\bibinfo {volume} {35}}, \bibinfo {pages} {933--938} (\bibinfo {year} {2018})}\BibitemShut {NoStop}%
\bibitem [{\citenamefont {Griffith} \emph {et~al.}(2016)\citenamefont {Griffith}, \citenamefont {Yu}, \citenamefont {Okawachi}, \citenamefont {Cardenas}, \citenamefont {Mohanty}, \citenamefont {Gaeta}, and \citenamefont {Lipson}}]{griffith2016coherent}%
  \BibitemOpen
  \bibfield  {author} {\bibinfo {author} {\bibfnamefont {A.~G.} \bibnamefont {Griffith}}, \bibinfo {author} {\bibfnamefont {M.}~\bibnamefont {Yu}}, \bibinfo {author} {\bibfnamefont {Y.}~\bibnamefont {Okawachi}}, \bibinfo {author} {\bibfnamefont {J.}~\bibnamefont {Cardenas}}, \bibinfo {author} {\bibfnamefont {A.}~\bibnamefont {Mohanty}}, \bibinfo {author} {\bibfnamefont {A.~L.} \bibnamefont {Gaeta}},  and \bibinfo {author} {\bibfnamefont {M.}~\bibnamefont {Lipson}}, }\bibfield  {title} {\enquote {\bibinfo {title} {{Coherent mid-infrared frequency combs in silicon-microresonators in the presence of Raman effects}},} }\href@noop {} {\bibfield  {journal} {\bibinfo  {journal} {Opt. Express} }\textbf {\bibinfo {volume} {24}}, \bibinfo {pages} {13044--13050} (\bibinfo {year} {2016})}\BibitemShut {NoStop}%
\bibitem [{\citenamefont {Zhang} \emph {et~al.}(2022{\natexlab{a}})\citenamefont {Zhang}, \citenamefont {Zhong}, \citenamefont {Zhou}, and \citenamefont {Tsang}}]{zhang2022broadband}%
  \BibitemOpen
  \bibfield  {author} {\bibinfo {author} {\bibfnamefont {Y.}~\bibnamefont {Zhang}}, \bibinfo {author} {\bibfnamefont {K.}~\bibnamefont {Zhong}}, \bibinfo {author} {\bibfnamefont {X.}~\bibnamefont {Zhou}},  and \bibinfo {author} {\bibfnamefont {H.~K.} \bibnamefont {Tsang}}, }\bibfield  {title} {\enquote {\bibinfo {title} {{Broadband high-Q multimode silicon concentric racetrack resonators for widely tunable Raman lasers}},} }\href@noop {} {\bibfield  {journal} {\bibinfo  {journal} {Nat. Commun.} }\textbf {\bibinfo {volume} {13}}, \bibinfo {pages} {3534} (\bibinfo {year} {2022}{\natexlab{a}})}\BibitemShut {NoStop}%
\bibitem [{\citenamefont {Zhang} \emph {et~al.}(2021)\citenamefont {Zhang}, \citenamefont {Zhong}, and \citenamefont {Tsang}}]{zhang2021raman}%
  \BibitemOpen
  \bibfield  {author} {\bibinfo {author} {\bibfnamefont {Y.}~\bibnamefont {Zhang}}, \bibinfo {author} {\bibfnamefont {K.}~\bibnamefont {Zhong}},  and \bibinfo {author} {\bibfnamefont {H.~K.} \bibnamefont {Tsang}}, }\bibfield  {title} {\enquote {\bibinfo {title} {{Raman lasing in multimode silicon racetrack resonators}},} }\href@noop {} {\bibfield  {journal} {\bibinfo  {journal} {Laser Photonics Rev.} }\textbf {\bibinfo {volume} {15}}, \bibinfo {pages} {2000336} (\bibinfo {year} {2021})}\BibitemShut {NoStop}%
\bibitem [{\citenamefont {Liu} \emph {et~al.}(2017)\citenamefont {Liu}, \citenamefont {Sun}, \citenamefont {Xiong}, \citenamefont {Wang}, \citenamefont {Wang}, \citenamefont {Han}, \citenamefont {Hao}, \citenamefont {Li}, \citenamefont {Luo}, \citenamefont {Yan} \emph {et~al.}}]{liu2017integrated}%
  \BibitemOpen
  \bibfield  {author} {\bibinfo {author} {\bibfnamefont {X.}~\bibnamefont {Liu}}, \bibinfo {author} {\bibfnamefont {C.}~\bibnamefont {Sun}}, \bibinfo {author} {\bibfnamefont {B.}~\bibnamefont {Xiong}}, \bibinfo {author} {\bibfnamefont {L.}~\bibnamefont {Wang}}, \bibinfo {author} {\bibfnamefont {J.}~\bibnamefont {Wang}}, \bibinfo {author} {\bibfnamefont {Y.}~\bibnamefont {Han}}, \bibinfo {author} {\bibfnamefont {Z.}~\bibnamefont {Hao}}, \bibinfo {author} {\bibfnamefont {H.}~\bibnamefont {Li}}, \bibinfo {author} {\bibfnamefont {Y.}~\bibnamefont {Luo}}, \bibinfo {author} {\bibfnamefont {J.}~\bibnamefont {Yan}},  \emph {et~al.}, }\bibfield  {title} {\enquote {\bibinfo {title} {{Integrated continuous-wave aluminum nitride Raman laser}},} }\href@noop {} {\bibfield  {journal} {\bibinfo  {journal} {Optica} }\textbf {\bibinfo {volume} {4}}, \bibinfo {pages} {893--896} (\bibinfo {year} {2017})}\BibitemShut {NoStop}%
\bibitem [{\citenamefont {Liu} \emph {et~al.}(2018)\citenamefont {Liu}, \citenamefont {Sun}, \citenamefont {Xiong}, \citenamefont {Wang}, \citenamefont {Wang}, \citenamefont {Han}, \citenamefont {Hao}, \citenamefont {Li}, \citenamefont {Luo}, \citenamefont {Yan} \emph {et~al.}}]{liu2018integrated}%
  \BibitemOpen
  \bibfield  {author} {\bibinfo {author} {\bibfnamefont {X.}~\bibnamefont {Liu}}, \bibinfo {author} {\bibfnamefont {C.}~\bibnamefont {Sun}}, \bibinfo {author} {\bibfnamefont {B.}~\bibnamefont {Xiong}}, \bibinfo {author} {\bibfnamefont {L.}~\bibnamefont {Wang}}, \bibinfo {author} {\bibfnamefont {J.}~\bibnamefont {Wang}}, \bibinfo {author} {\bibfnamefont {Y.}~\bibnamefont {Han}}, \bibinfo {author} {\bibfnamefont {Z.}~\bibnamefont {Hao}}, \bibinfo {author} {\bibfnamefont {H.}~\bibnamefont {Li}}, \bibinfo {author} {\bibfnamefont {Y.}~\bibnamefont {Luo}}, \bibinfo {author} {\bibfnamefont {J.}~\bibnamefont {Yan}},  \emph {et~al.}, }\bibfield  {title} {\enquote {\bibinfo {title} {{Integrated high-Q crystalline AlN microresonators for broadband Kerr and Raman frequency combs}},} }\href@noop {} {\bibfield  {journal} {\bibinfo  {journal} {ACS Photonics} }\textbf {\bibinfo {volume} {5}}, \bibinfo {pages} {1943--1950} (\bibinfo {year} {2018})}\BibitemShut {NoStop}%
\bibitem [{\citenamefont {Yu} \emph {et~al.}(2020)\citenamefont {Yu}, \citenamefont {Okawachi}, \citenamefont {Cheng}, \citenamefont {Wang}, \citenamefont {Zhang}, \citenamefont {Gaeta}, and \citenamefont {Lon{\v{c}}ar}}]{yu2020raman}%
  \BibitemOpen
  \bibfield  {author} {\bibinfo {author} {\bibfnamefont {M.}~\bibnamefont {Yu}}, \bibinfo {author} {\bibfnamefont {Y.}~\bibnamefont {Okawachi}}, \bibinfo {author} {\bibfnamefont {R.}~\bibnamefont {Cheng}}, \bibinfo {author} {\bibfnamefont {C.}~\bibnamefont {Wang}}, \bibinfo {author} {\bibfnamefont {M.}~\bibnamefont {Zhang}}, \bibinfo {author} {\bibfnamefont {A.~L.} \bibnamefont {Gaeta}},  and \bibinfo {author} {\bibfnamefont {M.}~\bibnamefont {Lon{\v{c}}ar}}, }\bibfield  {title} {\enquote {\bibinfo {title} {{Raman lasing and soliton mode-locking in lithium niobate microresonators}},} }\href@noop {} {\bibfield  {journal} {\bibinfo  {journal} {Light Sci. Appl.} }\textbf {\bibinfo {volume} {9}}, \bibinfo {pages} {9} (\bibinfo {year} {2020})}\BibitemShut {NoStop}%
\bibitem [{\citenamefont {Li} \emph {et~al.}(2024)\citenamefont {Li}, \citenamefont {Wang}, \citenamefont {Afridi}, \citenamefont {Lu}, \citenamefont {Shi}, \citenamefont {Sun}, \citenamefont {Ou}, and \citenamefont {Li}}]{li2024efficient}%
  \BibitemOpen
  \bibfield  {author} {\bibinfo {author} {\bibfnamefont {J.}~\bibnamefont {Li}}, \bibinfo {author} {\bibfnamefont {R.}~\bibnamefont {Wang}}, \bibinfo {author} {\bibfnamefont {A.~A.} \bibnamefont {Afridi}}, \bibinfo {author} {\bibfnamefont {Y.}~\bibnamefont {Lu}}, \bibinfo {author} {\bibfnamefont {X.}~\bibnamefont {Shi}}, \bibinfo {author} {\bibfnamefont {W.}~\bibnamefont {Sun}}, \bibinfo {author} {\bibfnamefont {H.}~\bibnamefont {Ou}},  and \bibinfo {author} {\bibfnamefont {Q.}~\bibnamefont {Li}}, }\bibfield  {title} {\enquote {\bibinfo {title} {{Efficient Raman lasing and Raman--Kerr interaction in an integrated silicon carbide platform}},} }\href@noop {} {\bibfield  {journal} {\bibinfo  {journal} {ACS Photonics} }\textbf {\bibinfo {volume} {11}}, \bibinfo {pages} {795--800} (\bibinfo {year} {2024})}\BibitemShut {NoStop}%
\bibitem [{\citenamefont {Hausmann} \emph {et~al.}(2014)\citenamefont {Hausmann}, \citenamefont {Bulu}, \citenamefont {Venkataraman}, \citenamefont {Deotare}, and \citenamefont {Lon{\v{c}}ar}}]{hausmann2014diamond}%
  \BibitemOpen
  \bibfield  {author} {\bibinfo {author} {\bibfnamefont {B.}~\bibnamefont {Hausmann}}, \bibinfo {author} {\bibfnamefont {I.}~\bibnamefont {Bulu}}, \bibinfo {author} {\bibfnamefont {V.}~\bibnamefont {Venkataraman}}, \bibinfo {author} {\bibfnamefont {P.}~\bibnamefont {Deotare}},  and \bibinfo {author} {\bibfnamefont {M.}~\bibnamefont {Lon{\v{c}}ar}}, }\bibfield  {title} {\enquote {\bibinfo {title} {{Diamond nonlinear photonics}},} }\href@noop {} {\bibfield  {journal} {\bibinfo  {journal} {Nat. Photon.} }\textbf {\bibinfo {volume} {8}}, \bibinfo {pages} {369--374} (\bibinfo {year} {2014})}\BibitemShut {NoStop}%
\bibitem [{\citenamefont {Choi} \emph {et~al.}(2019)\citenamefont {Choi}, \citenamefont {Chen}, \citenamefont {Du}, \citenamefont {Zeto}, and \citenamefont {Armani}}]{choi2019low}%
  \BibitemOpen
  \bibfield  {author} {\bibinfo {author} {\bibfnamefont {H.}~\bibnamefont {Choi}}, \bibinfo {author} {\bibfnamefont {D.}~\bibnamefont {Chen}}, \bibinfo {author} {\bibfnamefont {F.}~\bibnamefont {Du}}, \bibinfo {author} {\bibfnamefont {R.}~\bibnamefont {Zeto}},  and \bibinfo {author} {\bibfnamefont {A.}~\bibnamefont {Armani}}, }\bibfield  {title} {\enquote {\bibinfo {title} {{Low threshold anti-Stokes Raman laser on-chip}},} }\href@noop {} {\bibfield  {journal} {\bibinfo  {journal} {Photonics Res.} }\textbf {\bibinfo {volume} {7}}, \bibinfo {pages} {926--932} (\bibinfo {year} {2019})}\BibitemShut {NoStop}%
\bibitem [{\citenamefont {Xia} \emph {et~al.}(2022)\citenamefont {Xia}, \citenamefont {Huang}, \citenamefont {Zhang}, \citenamefont {Zeng}, \citenamefont {Zhao}, \citenamefont {Yang}, \citenamefont {Sun}, \citenamefont {Luo}, \citenamefont {Hu}, \citenamefont {Liu} \emph {et~al.}}]{xia2022engineered}%
  \BibitemOpen
  \bibfield  {author} {\bibinfo {author} {\bibfnamefont {D.}~\bibnamefont {Xia}}, \bibinfo {author} {\bibfnamefont {Y.}~\bibnamefont {Huang}}, \bibinfo {author} {\bibfnamefont {B.}~\bibnamefont {Zhang}}, \bibinfo {author} {\bibfnamefont {P.}~\bibnamefont {Zeng}}, \bibinfo {author} {\bibfnamefont {J.}~\bibnamefont {Zhao}}, \bibinfo {author} {\bibfnamefont {Z.}~\bibnamefont {Yang}}, \bibinfo {author} {\bibfnamefont {S.}~\bibnamefont {Sun}}, \bibinfo {author} {\bibfnamefont {L.}~\bibnamefont {Luo}}, \bibinfo {author} {\bibfnamefont {G.}~\bibnamefont {Hu}}, \bibinfo {author} {\bibfnamefont {D.}~\bibnamefont {Liu}},  \emph {et~al.}, }\bibfield  {title} {\enquote {\bibinfo {title} {{Engineered Raman lasing in photonic integrated chalcogenide microresonators}},} }\href@noop {} {\bibfield  {journal} {\bibinfo  {journal} {Laser Photonics Rev.} }\textbf {\bibinfo {volume} {16}}, \bibinfo {pages} {2100443} (\bibinfo {year} {2022})}\BibitemShut {NoStop}%
\bibitem [{\citenamefont {Okawachi} \emph {et~al.}(2011)\citenamefont {Okawachi}, \citenamefont {Saha}, \citenamefont {Levy}, \citenamefont {Wen}, \citenamefont {Lipson}, and \citenamefont {Gaeta}}]{okawachi2011octave}%
  \BibitemOpen
  \bibfield  {author} {\bibinfo {author} {\bibfnamefont {Y.}~\bibnamefont {Okawachi}}, \bibinfo {author} {\bibfnamefont {K.}~\bibnamefont {Saha}}, \bibinfo {author} {\bibfnamefont {J.~S.} \bibnamefont {Levy}}, \bibinfo {author} {\bibfnamefont {Y.~H.} \bibnamefont {Wen}}, \bibinfo {author} {\bibfnamefont {M.}~\bibnamefont {Lipson}},  and \bibinfo {author} {\bibfnamefont {A.~L.} \bibnamefont {Gaeta}}, }\bibfield  {title} {\enquote {\bibinfo {title} {{Octave-spanning frequency comb generation in a silicon nitride chip}},} }\href@noop {} {\bibfield  {journal} {\bibinfo  {journal} {Opt. Lett.} }\textbf {\bibinfo {volume} {36}}, \bibinfo {pages} {3398--3400} (\bibinfo {year} {2011})}\BibitemShut {NoStop}%
\bibitem [{\citenamefont {Brasch} \emph {et~al.}(2016)\citenamefont {Brasch}, \citenamefont {Geiselmann}, \citenamefont {Herr}, \citenamefont {Lihachev}, \citenamefont {Pfeiffer}, \citenamefont {Gorodetsky}, and \citenamefont {Kippenberg}}]{brasch2016photonic}%
  \BibitemOpen
  \bibfield  {author} {\bibinfo {author} {\bibfnamefont {V.}~\bibnamefont {Brasch}}, \bibinfo {author} {\bibfnamefont {M.}~\bibnamefont {Geiselmann}}, \bibinfo {author} {\bibfnamefont {T.}~\bibnamefont {Herr}}, \bibinfo {author} {\bibfnamefont {G.}~\bibnamefont {Lihachev}}, \bibinfo {author} {\bibfnamefont {M.~H.} \bibnamefont {Pfeiffer}}, \bibinfo {author} {\bibfnamefont {M.~L.} \bibnamefont {Gorodetsky}},  and \bibinfo {author} {\bibfnamefont {T.~J.} \bibnamefont {Kippenberg}}, }\bibfield  {title} {\enquote {\bibinfo {title} {{Photonic chip--based optical frequency comb using soliton Cherenkov radiation}},} }\href@noop {} {\bibfield  {journal} {\bibinfo  {journal} {Science} }\textbf {\bibinfo {volume} {351}}, \bibinfo {pages} {357--360} (\bibinfo {year} {2016})}\BibitemShut {NoStop}%
\bibitem [{\citenamefont {Rebolledo-Salgado} \emph {et~al.}(2022)\citenamefont {Rebolledo-Salgado}, \citenamefont {Ye}, \citenamefont {Christensen}, \citenamefont {Lei}, \citenamefont {Twayana}, \citenamefont {Schr{\"o}der}, \citenamefont {Zelan}, and \citenamefont {Torres-Company}}]{rebolledo2022coherent}%
  \BibitemOpen
  \bibfield  {author} {\bibinfo {author} {\bibfnamefont {I.}~\bibnamefont {Rebolledo-Salgado}}, \bibinfo {author} {\bibfnamefont {Z.}~\bibnamefont {Ye}}, \bibinfo {author} {\bibfnamefont {S.}~\bibnamefont {Christensen}}, \bibinfo {author} {\bibfnamefont {F.}~\bibnamefont {Lei}}, \bibinfo {author} {\bibfnamefont {K.}~\bibnamefont {Twayana}}, \bibinfo {author} {\bibfnamefont {J.}~\bibnamefont {Schr{\"o}der}}, \bibinfo {author} {\bibfnamefont {M.}~\bibnamefont {Zelan}},  and \bibinfo {author} {\bibfnamefont {V.}~\bibnamefont {Torres-Company}}, }\bibfield  {title} {\enquote {\bibinfo {title} {{Coherent supercontinuum generation in all-normal dispersion Si3N4 waveguides}},} }\href@noop {} {\bibfield  {journal} {\bibinfo  {journal} {Opt. Express} }\textbf {\bibinfo {volume} {30}}, \bibinfo {pages} {8641--8651} (\bibinfo {year} {2022})}\BibitemShut {NoStop}%
\bibitem [{\citenamefont {Rahim} \emph {et~al.}(2017)\citenamefont {Rahim}, \citenamefont {Ryckeboer}, \citenamefont {Subramanian}, \citenamefont {Clemmen}, \citenamefont {Kuyken}, \citenamefont {Dhakal}, \citenamefont {Raza}, \citenamefont {Hermans}, \citenamefont {Muneeb}, \citenamefont {Dhoore} \emph {et~al.}}]{rahim2017expanding}%
  \BibitemOpen
  \bibfield  {author} {\bibinfo {author} {\bibfnamefont {A.}~\bibnamefont {Rahim}}, \bibinfo {author} {\bibfnamefont {E.}~\bibnamefont {Ryckeboer}}, \bibinfo {author} {\bibfnamefont {A.~Z.} \bibnamefont {Subramanian}}, \bibinfo {author} {\bibfnamefont {S.}~\bibnamefont {Clemmen}}, \bibinfo {author} {\bibfnamefont {B.}~\bibnamefont {Kuyken}}, \bibinfo {author} {\bibfnamefont {A.}~\bibnamefont {Dhakal}}, \bibinfo {author} {\bibfnamefont {A.}~\bibnamefont {Raza}}, \bibinfo {author} {\bibfnamefont {A.}~\bibnamefont {Hermans}}, \bibinfo {author} {\bibfnamefont {M.}~\bibnamefont {Muneeb}}, \bibinfo {author} {\bibfnamefont {S.}~\bibnamefont {Dhoore}},  \emph {et~al.}, }\bibfield  {title} {\enquote {\bibinfo {title} {{Expanding the silicon photonics portfolio with silicon nitride photonic integrated circuits}},} }\href@noop {} {\bibfield  {journal} {\bibinfo  {journal} {J. Light. Technol.} }\textbf {\bibinfo {volume} {35}}, \bibinfo {pages} {639--649} (\bibinfo {year} {2017})}\BibitemShut {NoStop}%
\bibitem [{\citenamefont {Zhang} \emph {et~al.}(2024{\natexlab{a}})\citenamefont {Zhang}, \citenamefont {Bi}, \citenamefont {Harder}, \citenamefont {Ohletz}, \citenamefont {Gannott}, \citenamefont {Gumann}, \citenamefont {Butzen}, \citenamefont {Zhang}, and \citenamefont {Del'Haye}}]{zhang2024low}%
  \BibitemOpen
  \bibfield  {author} {\bibinfo {author} {\bibfnamefont {S.}~\bibnamefont {Zhang}}, \bibinfo {author} {\bibfnamefont {T.}~\bibnamefont {Bi}}, \bibinfo {author} {\bibfnamefont {I.}~\bibnamefont {Harder}}, \bibinfo {author} {\bibfnamefont {O.}~\bibnamefont {Ohletz}}, \bibinfo {author} {\bibfnamefont {F.}~\bibnamefont {Gannott}}, \bibinfo {author} {\bibfnamefont {A.}~\bibnamefont {Gumann}}, \bibinfo {author} {\bibfnamefont {E.}~\bibnamefont {Butzen}}, \bibinfo {author} {\bibfnamefont {Y.}~\bibnamefont {Zhang}},  and \bibinfo {author} {\bibfnamefont {P.}~\bibnamefont {Del'Haye}}, }\bibfield  {title} {\enquote {\bibinfo {title} {{Low-Temperature Sputtered Ultralow-Loss Silicon Nitride for Hybrid Photonic Integration}},} }\href@noop {} {\bibfield  {journal} {\bibinfo  {journal} {Laser Photonics Rev.} }\textbf {\bibinfo {volume} {18}}, \bibinfo {pages} {2300642} (\bibinfo {year} {2024}{\natexlab{a}})}\BibitemShut {NoStop}%
\bibitem [{\citenamefont {Chiles} \emph {et~al.}(2018)\citenamefont {Chiles}, \citenamefont {Nader}, \citenamefont {Hickstein}, \citenamefont {Yu}, \citenamefont {Briles}, \citenamefont {Carlson}, \citenamefont {Jung}, \citenamefont {Shainline}, \citenamefont {Diddams}, \citenamefont {Papp} \emph {et~al.}}]{chiles2018deuterated}%
  \BibitemOpen
  \bibfield  {author} {\bibinfo {author} {\bibfnamefont {J.}~\bibnamefont {Chiles}}, \bibinfo {author} {\bibfnamefont {N.}~\bibnamefont {Nader}}, \bibinfo {author} {\bibfnamefont {D.~D.} \bibnamefont {Hickstein}}, \bibinfo {author} {\bibfnamefont {S.~P.} \bibnamefont {Yu}}, \bibinfo {author} {\bibfnamefont {T.~C.} \bibnamefont {Briles}}, \bibinfo {author} {\bibfnamefont {D.}~\bibnamefont {Carlson}}, \bibinfo {author} {\bibfnamefont {H.}~\bibnamefont {Jung}}, \bibinfo {author} {\bibfnamefont {J.~M.} \bibnamefont {Shainline}}, \bibinfo {author} {\bibfnamefont {S.}~\bibnamefont {Diddams}}, \bibinfo {author} {\bibfnamefont {S.~B.} \bibnamefont {Papp}},  \emph {et~al.}, }\bibfield  {title} {\enquote {\bibinfo {title} {{Deuterated silicon nitride photonic devices for broadband optical frequency comb generation}},} }\href@noop {} {\bibfield  {journal} {\bibinfo  {journal} {Opt. Lett.} }\textbf {\bibinfo {volume} {43}}, \bibinfo {pages} {1527--1530} (\bibinfo {year} {2018})}\BibitemShut {NoStop}%
\bibitem [{\citenamefont {Zhang} \emph {et~al.}(2025)\citenamefont {Zhang}, \citenamefont {Zhang}, \citenamefont {Ghosh}, \citenamefont {Pal}, \citenamefont {Ghalanos~N}, \citenamefont {Bi}, \citenamefont {Yan}, \citenamefont {Zhang}, \citenamefont {Zhuang}, \citenamefont {Hill}, and \citenamefont {Del’Haye}}]{yaojing_SSB}%
  \BibitemOpen
  \bibfield  {author} {\bibinfo {author} {\bibfnamefont {Y.}~\bibnamefont {Zhang}}, \bibinfo {author} {\bibfnamefont {S.}~\bibnamefont {Zhang}}, \bibinfo {author} {\bibfnamefont {A.}~\bibnamefont {Ghosh}}, \bibinfo {author} {\bibfnamefont {A.}~\bibnamefont {Pal}}, \bibinfo {author} {\bibfnamefont {G.}~\bibnamefont {Ghalanos~N}}, \bibinfo {author} {\bibfnamefont {T.}~\bibnamefont {Bi}}, \bibinfo {author} {\bibfnamefont {H.}~\bibnamefont {Yan}}, \bibinfo {author} {\bibfnamefont {H.}~\bibnamefont {Zhang}}, \bibinfo {author} {\bibfnamefont {Y.}~\bibnamefont {Zhuang}}, \bibinfo {author} {\bibfnamefont {L.}~\bibnamefont {Hill}},  and \bibinfo {author} {\bibfnamefont {P.}~\bibnamefont {Del’Haye}}, }\bibfield  {title} {\enquote {\bibinfo {title} {Integrated optical switches based on kerr symmetry breaking in microresonators},} }\href@noop {} {\bibfield  {journal} {\bibinfo  {journal} {Photonics Res.} }\textbf {\bibinfo {volume} {13}}, \bibinfo {pages} {360--366} (\bibinfo {year} {2025})}\BibitemShut {NoStop}%
\bibitem [{\citenamefont {Karpov} \emph {et~al.}(2016)\citenamefont {Karpov}, \citenamefont {Guo}, \citenamefont {Kordts}, \citenamefont {Brasch}, \citenamefont {Pfeiffer}, \citenamefont {Zervas}, \citenamefont {Geiselmann}, and \citenamefont {Kippenberg}}]{karpov2016raman}%
  \BibitemOpen
  \bibfield  {author} {\bibinfo {author} {\bibfnamefont {M.}~\bibnamefont {Karpov}}, \bibinfo {author} {\bibfnamefont {H.}~\bibnamefont {Guo}}, \bibinfo {author} {\bibfnamefont {A.}~\bibnamefont {Kordts}}, \bibinfo {author} {\bibfnamefont {V.}~\bibnamefont {Brasch}}, \bibinfo {author} {\bibfnamefont {M.~H.} \bibnamefont {Pfeiffer}}, \bibinfo {author} {\bibfnamefont {M.}~\bibnamefont {Zervas}}, \bibinfo {author} {\bibfnamefont {M.}~\bibnamefont {Geiselmann}},  and \bibinfo {author} {\bibfnamefont {T.~J.} \bibnamefont {Kippenberg}}, }\bibfield  {title} {\enquote {\bibinfo {title} {{Raman self-frequency shift of dissipative Kerr solitons in an optical microresonator}},} }\href@noop {} {\bibfield  {journal} {\bibinfo  {journal} {Phys. Rev. Lett.} }\textbf {\bibinfo {volume} {116}}, \bibinfo {pages} {103902} (\bibinfo {year} {2016})}\BibitemShut {NoStop}%
\bibitem [{\citenamefont {Zhang} \emph {et~al.}(2023)\citenamefont {Zhang}, \citenamefont {Zhang}, \citenamefont {Bi}, and \citenamefont {Del’Haye}}]{zhang2023geometry}%
  \BibitemOpen
  \bibfield  {author} {\bibinfo {author} {\bibfnamefont {Y.}~\bibnamefont {Zhang}}, \bibinfo {author} {\bibfnamefont {S.}~\bibnamefont {Zhang}}, \bibinfo {author} {\bibfnamefont {T.}~\bibnamefont {Bi}},  and \bibinfo {author} {\bibfnamefont {P.}~\bibnamefont {Del’Haye}}, }\bibfield  {title} {\enquote {\bibinfo {title} {{Geometry optimization for dark soliton combs in thin multimode silicon nitride microresonators}},} }\href@noop {} {\bibfield  {journal} {\bibinfo  {journal} {Opt. Express} }\textbf {\bibinfo {volume} {31}}, \bibinfo {pages} {41420--41427} (\bibinfo {year} {2023})}\BibitemShut {NoStop}%
\bibitem [{\citenamefont {Strekalov} and \citenamefont {Yu}(2009)}]{strekalov2009generation}%
  \BibitemOpen
  \bibfield  {author} {\bibinfo {author} {\bibfnamefont {D.~V.} \bibnamefont {Strekalov}} and \bibinfo {author} {\bibfnamefont {N.}~\bibnamefont {Yu}}, }\bibfield  {title} {\enquote {\bibinfo {title} {{Generation of optical combs in a whispering gallery mode resonator from a bichromatic pump}},} }\href@noop {} {\bibfield  {journal} {\bibinfo  {journal} {Phys. Rev. A} }\textbf {\bibinfo {volume} {79}}, \bibinfo {pages} {041805} (\bibinfo {year} {2009})}\BibitemShut {NoStop}%
\bibitem [{\citenamefont {Zhang} \emph {et~al.}(2020)\citenamefont {Zhang}, \citenamefont {Silver}, \citenamefont {Bi}, and \citenamefont {Del’Haye}}]{zhang2020spectral}%
  \BibitemOpen
  \bibfield  {author} {\bibinfo {author} {\bibfnamefont {S.}~\bibnamefont {Zhang}}, \bibinfo {author} {\bibfnamefont {J.~M.} \bibnamefont {Silver}}, \bibinfo {author} {\bibfnamefont {T.}~\bibnamefont {Bi}},  and \bibinfo {author} {\bibfnamefont {P.}~\bibnamefont {Del’Haye}}, }\bibfield  {title} {\enquote {\bibinfo {title} {{Spectral extension and synchronization of microcombs in a single microresonator}},} }\href@noop {} {\bibfield  {journal} {\bibinfo  {journal} {Nat. Commun.} }\textbf {\bibinfo {volume} {11}}, \bibinfo {pages} {6384} (\bibinfo {year} {2020})}\BibitemShut {NoStop}%
\bibitem [{\citenamefont {Zhang} \emph {et~al.}(2022{\natexlab{b}})\citenamefont {Zhang}, \citenamefont {Bi}, \citenamefont {Ghalanos}, \citenamefont {Moroney}, \citenamefont {Del~Bino}, and \citenamefont {Del’Haye}}]{zhang2022dark}%
  \BibitemOpen
  \bibfield  {author} {\bibinfo {author} {\bibfnamefont {S.}~\bibnamefont {Zhang}}, \bibinfo {author} {\bibfnamefont {T.}~\bibnamefont {Bi}}, \bibinfo {author} {\bibfnamefont {G.~N.} \bibnamefont {Ghalanos}}, \bibinfo {author} {\bibfnamefont {N.~P.} \bibnamefont {Moroney}}, \bibinfo {author} {\bibfnamefont {L.}~\bibnamefont {Del~Bino}},  and \bibinfo {author} {\bibfnamefont {P.}~\bibnamefont {Del’Haye}}, }\bibfield  {title} {\enquote {\bibinfo {title} {{Dark-bright soliton bound states in a microresonator}},} }\href@noop {} {\bibfield  {journal} {\bibinfo  {journal} {Phys. Rev. Lett.} }\textbf {\bibinfo {volume} {128}}, \bibinfo {pages} {033901} (\bibinfo {year} {2022}{\natexlab{b}})}\BibitemShut {NoStop}%
\bibitem [{\citenamefont {Kippenberg} \emph {et~al.}(2004)\citenamefont {Kippenberg}, \citenamefont {Spillane}, \citenamefont {Min}, and \citenamefont {Vahala}}]{kippenberg2004theoretical}%
  \BibitemOpen
  \bibfield  {author} {\bibinfo {author} {\bibfnamefont {T.~J.} \bibnamefont {Kippenberg}}, \bibinfo {author} {\bibfnamefont {S.~M.} \bibnamefont {Spillane}}, \bibinfo {author} {\bibfnamefont {B.}~\bibnamefont {Min}},  and \bibinfo {author} {\bibfnamefont {K.~J.} \bibnamefont {Vahala}}, }\bibfield  {title} {\enquote {\bibinfo {title} {{Theoretical and experimental study of stimulated and cascaded Raman scattering in ultrahigh-Q optical microcavities}},} }\href@noop {} {\bibfield  {journal} {\bibinfo  {journal} {IEEE J. Sel. Topics Quantum Electron.} }\textbf {\bibinfo {volume} {10}}, \bibinfo {pages} {1219--1228} (\bibinfo {year} {2004})}\BibitemShut {NoStop}%
\bibitem [{\citenamefont {Zhang} \emph {et~al.}(2024{\natexlab{b}})\citenamefont {Zhang}, \citenamefont {Bi}, and \citenamefont {Del’Haye}}]{zhang2024fly}%
  \BibitemOpen
  \bibfield  {author} {\bibinfo {author} {\bibfnamefont {S.}~\bibnamefont {Zhang}}, \bibinfo {author} {\bibfnamefont {T.}~\bibnamefont {Bi}},  and \bibinfo {author} {\bibfnamefont {P.}~\bibnamefont {Del’Haye}}, }\bibfield  {title} {\enquote {\bibinfo {title} {{On-the-fly precision spectroscopy with a dual-modulated tunable diode laser and Hz-level referencing to a cavity}},} }\href@noop {} {\bibfield  {journal} {\bibinfo  {journal} {Adv. Photonics} }\textbf {\bibinfo {volume} {6}}, \bibinfo {pages} {046003--046003} (\bibinfo {year} {2024}{\natexlab{b}})}\BibitemShut {NoStop}%
\bibitem [{\citenamefont {Pal} \emph {et~al.}(2023)\citenamefont {Pal}, \citenamefont {Ghosh}, \citenamefont {Zhang}, \citenamefont {Bi}, and \citenamefont {Del’Haye}}]{pal2023machine}%
  \BibitemOpen
  \bibfield  {author} {\bibinfo {author} {\bibfnamefont {A.}~\bibnamefont {Pal}}, \bibinfo {author} {\bibfnamefont {A.}~\bibnamefont {Ghosh}}, \bibinfo {author} {\bibfnamefont {S.}~\bibnamefont {Zhang}}, \bibinfo {author} {\bibfnamefont {T.}~\bibnamefont {Bi}},  and \bibinfo {author} {\bibfnamefont {P.}~\bibnamefont {Del’Haye}}, }\bibfield  {title} {\enquote {\bibinfo {title} {{Machine learning assisted inverse design of microresonators}},} }\href@noop {} {\bibfield  {journal} {\bibinfo  {journal} {Opt. Express} }\textbf {\bibinfo {volume} {31}}, \bibinfo {pages} {8020--8028} (\bibinfo {year} {2023})}\BibitemShut {NoStop}%
\end{thebibliography}%
\onecolumngrid

\appendix
\clearpage

\section{\large{Appendix A: Quality factor and dispersion characterization}}
\label{A}
\begin{figure*}[h!]
\includegraphics[width=0.6\columnwidth]{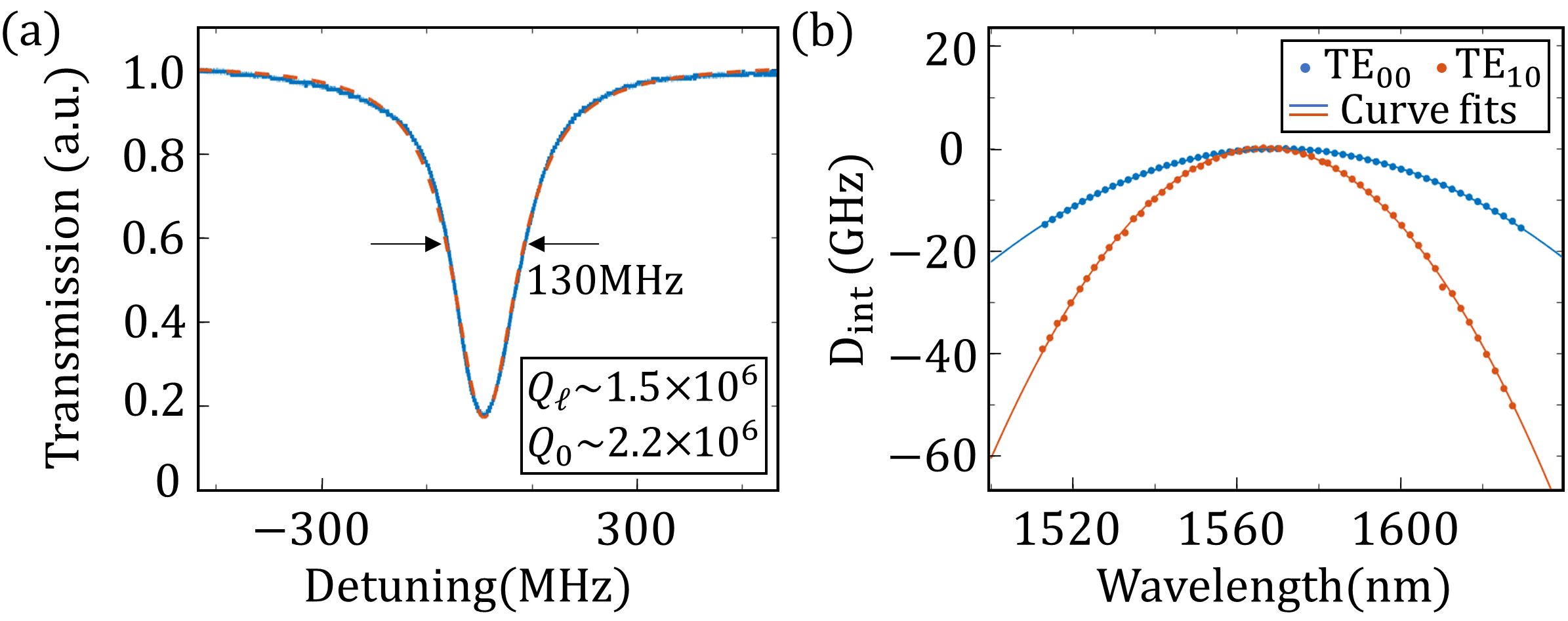}

\caption {\textit{Resonance linewidth and dispersion characterization of the Si$_3$N$_4$ ring-resonator.} (a) Transmission profile around $1570~$nm showing the fundamental mode's linewidth of around $130~$MHz. The loaded \textit{Q} ($\textit{Q}_\ell$) is $1.5\times10^{6}$ and the intrinsic \textit{Q} ($\textit{Q}_\text{0}$) is $2.2\times10^6$.
(b) Integrated dispersion for the fundamental mode (TE$_{00}$) and the higher-order mode (TE$_{10}$) families.}
\label{Appendix_A}
\end{figure*}

The transmission spectrum of a fundamental mode (TE$_{00}$) at 1570 nm with an intrinsic \textit{Q}-factor of 2.2 million is shown in Fig.\ref{Appendix_A}(d). High-precision tunable diode laser spectroscopy based on dual radio-frequency modulation is used to characterize the fabricated Si$_3$N$_4$ microresonators~\cite{zhang2024fly}. The experimentally obtained integrated dispersion profiles ($D\mathrm{_{int}}$)~\cite{pal2023machine} in Fig.~\ref{Appendix_A}(e) indicate normal dispersions for both the fundamental (TE$_{00}$) mode and the higher order (TE$_{10}$) mode. 

\end{document}